\documentclass[published]{JHEP3} 

\JHEP{00(2002)000}

\JHEPspecialurl{http://jhep.sissa.it/JOURNAL/JHEP3.tar.gz}

\usepackage{epsfig,multicol}

\newcommand\fverb{\setbox\pippobox=\hbox\bgroup\verb}
\newcommand\fverbdo{\egroup\medskip\noindent%
            \fbox{\unhbox\pippobox}\ }
\newcommand\fverbit{\egroup\item[\fbox{\unhbox\pippobox}]}
\newbox\pippobox

\title{4D gravity localized in non $Z_2$--symmetric thick branes}

\author{Nandinii Barbosa--Cendejas~ and Alfredo Herrera--Aguilar\\
Instituto de F\'{\i}sica y Matem\'{a}ticas, Universidad Michoacana
de San Nicol\'as de Hidalgo. Edificio C-3, Cd. Universitaria.
C. P. 58040 Morelia, Michoac\'{a}n, M\'{e}xico.\\
E-mail: \email{nandinii@ifm.umich.mx}, \email{herrera@ifm.umich.mx}}
\received{October 18, 2005}          
\revised{October 24, 2005}
\accepted{October 25, 2005}          
\abstract{We present a comparative analysis of localization of 4D
gravity on a non $Z_2$--symmetric scalar thick brane in both a
5--dimensional Riemannian space time and a pure geometric Weyl
integrable manifold in which variations in the length of vectors
during parallel transport are allowed and a geometric scalar field
is involved in its formulation. This work was mainly motivated by
the hypothesis which claims that Weyl geometries mimic quantum
behaviour classically. We start by obtaining a classical
4--dimensional Poincar\'e invariant thick brane solution which does
not respect $Z_2$--symmetry along the (non--)compact extra
dimension. This field configuration reproduces the $Z_2$--symmetric
solutions previously found in the literature, in both the Riemann
and the Weyl frames, when the parameter $k_1=1$. The scalar energy
density of our field configuration represents several series of
thick branes with positive and negative energy densities centered at
$y_0$. Thus, our field configurations can be compared with the
standard Randall--Sundrum thin brane case. The only qualitative
difference we have encountered when comparing both frames is that
the scalar curvature of the Riemannian manifold turns out to be
singular for the found solution, whereas its Weylian counterpart
presents a regular behaviour. By studying the transverse traceless
modes of the fluctuations of the classical backgrounds, we recast
their equations into a Sch\"odinger's equation form with a volcano
potential of finite bottom (in both frames). By solving the
Sch\"odinger equation for the massless zero mode $m^2=0$ we obtain a
single bound state which represents a stable 4--dimensional graviton
in both frames. We also get a continuum gapless spectrum of KK
states with positive $m^2>0$ that are suppressed at $y_0$, turning
into continuum plane wave modes as $y$ approaches spatial infinity.
We show that for the considered solution to our setup, the potential
is always bounded and cannot adopt the form of a well with infinite
walls; thus, we do not get a discrete spectrum of KK states, and we
conclude that the claim that Weylian structures mimic, classically,
quantum behaviour does not constitute a generic feature of these
geometric manifolds.}


\begin{document}


\section{Introduction}
During last years it has shown an increasing interest in space times
with large extra dimensions since in these models gravity propagates
in all dimensions while matter is confined to a 4--dimensional
submanifold (a 3--brane) with no contradiction with present time
gravitational experiments \cite{rubakov}--\cite{gog}. In the
framework of brane scenarios in 5--dimensional space time it has
been shown a path towards the solution of some relevant problems of
high--energy physics such as the cosmological constant, dark matter,
non--locality and the mass hierarchy problem. In particular, it was
discovered that in such brane scenarios 4--dimensional gravity can
be realized consistently and we can live in $4+1$ non--compact
dimensions in perfect compatibility with experimental gravity
\cite{rs}--\cite{lr}. Since then, several generalizations of these
scenarios have been constructed with the aid of thick branes
\cite{dewolfe}--\cite{varios}. In this paper we present a
comparative study on the physics of a particular thick brane
solution made out of scalar matter in two different manifolds: a
Riemannian space time and a purely geometric manifold endowed with
Weyl structure. Thus, apart from the traditional Riemannian
treatment of the problem, where we have gravity coupled to a scalar
field, we shall consider a non--Riemannian generalization of
5--dimensional Kaluza--Klein theory by replacing the Riemannian
structure of the 5--dimensional manifold by a Weyl integrable
geometry which allows for variations in the length of vectors during
parallel transport and involves a geometric scalar field in its
formulation. It has been shown that in certain cases Weylian
structures mimic classically quantum behaviour \cite{quantbeh}. It
is interesting to see whether or not we can obtain such a behaviour
in our setup.

Thus, we begin by studying a 5--dimensional Weyl gravity model in
which branes (thick domain walls) can be obtained naturally without
introducing them by hand in the action of the theory. We further
implement the conformal technique to obtain a classical solution
that respects 4--dimensional Poincar\'e invariance and represents a
localized function with no reflection $Z_2$--symmetry and allows for
both compact and non--compact manifolds in the extra dimension. By
looking at the energy density of the scalar field of this solution
we interpret the field configuration as a set of non
$Z_2$--symmetric thick branes. We investigate as well the behaviour
of the curvature scalar and make an analysis of the fluctuations of
the metric around the classical background solution in order to know
whether 4--dimensional gravity can be described in our setup. It
turns out that this is the case since the quantum mechanical problem
with a potential well which vanishes asymptotically for the
transverse traceless sector of the fluctuations of the metric yields
a continuum spectrum of KK--states with a zero mode that corresponds
to the normalizable, stable 4--dimensional graviton.

Let us consider a non--Riemannian generalization of Kaluza--Klein
theory, namely, the pure geometrical 5--dimensional action
\begin{equation}
\label{action} S_5^W =\int_{M_5^W}\frac{d^5x\sqrt{|g|}}{16\pi
G_5}e^{\frac{3}{2}\omega}\left[R+3\tilde{\xi}\left(\nabla\omega\right)^2+6U(\omega)\right],
\end{equation}
where $M_5^W$ is a 5--dimensional Weyl integrable manifold
determined by $(g_{MN},\omega)$, $g_{MN}$ is the metric
($M,N=0,1,2,3,5$) and $\omega$ is a scalar function. In such
manifolds the Ricci tensor reads
$R_{MN}=\Gamma_{MN,A}^A-\Gamma_{AM,N}^A+\Gamma_{MN}^P\Gamma_{PQ}^Q-\Gamma_{MQ}^P\Gamma_{NP}^Q$,
where $\Gamma_{MN}^C=\{_{MN}^{\;C}\}-\frac{1}{2}
\left(\omega_{,M}\delta_N^C+\omega_{,N}\delta_M^C-g_{MN}\omega^{,C}\right)$
are the affine connections and $\{_{MN}^{\;C}\}$ are the Christoffel
symbols, the parameter $\tilde{\xi}$ is an arbitrary coupling
constant, and $U(\omega)$ is a self--interaction potential for the
scalar field. This action is of pure geometrical nature since the
scalar field that couples to gravity is precisely the scalar
function $\omega$ that enters in the definition of the affine
connections of the Weyl manifold and, thus, cannot be discarded in
principle from our consideration. Weyl integrable manifolds are
invariant under the so--called Weyl rescalings
\begin{eqnarray}
\label{weylrescalings} g_{MN}\rightarrow\Omega^{-2}g_{MN},\qquad
\omega\rightarrow\omega+\ln\Omega^2,\qquad
\tilde{\xi}\rightarrow\tilde{\xi}/\left(1+\partial_\omega\ln\Omega^2\right)^2,
\end{eqnarray} where $\Omega^2$ is a smooth function on $M_5^W$.
In general, this invariance is broken by the self--interaction
potential $U(\omega)$. From the relations (\ref{weylrescalings}) it
follows that the transformation $U\rightarrow\Omega^2U$ preserves
such an invariance. Thus, the potential $U(w)=\lambda e^{\omega}$,
where $\lambda$ is a coupling constant, preserves the scale
invariance of the Weyl action (\ref{action}). When the Weyl
invariance is broken, the scalar field is transformed into an
observable degree of freedom which models the thick branes.

In order to find 4--dimensional Poincar\'e invariant solutions of
the theory, we consider the following ansatz for the metric
\begin{equation}
\label{line} ds_5^2=e^{2A(y)}\eta_{mn}dx^m dx^n+dy^2,
\end{equation}
where $e^{2A(y)}$ is the warp factor, $m,n=0,1,2,3$  and $y$ labels
the extra dimension.

By taking into account (\ref{line}) we find the expressions for the
Ricci tensor and the scalar curvature in the Weyl frame
\begin{equation}
R_{mn}=-e^{2A}\left[A''+4(A')^2\right]\eta_{mn},  \qquad
R_{55}=-4\left[A''+(A')^2\right], \nonumber\\
\end{equation}
\begin{equation}
^5R=-4\left[2A''+5(A')^2\right],
\end{equation}
where the comma denotes derivatives with respect to the fifth
coordinate $y$. The 5--dimensional stress--energy tensor is given by
\begin{equation}
T_{MN}=\frac{1}{8\pi G_5}\left(R_{MN}-\frac{1}{2}g_{MN}R\right),
\end{equation}
thus, its 4--dimensional and pure 5--dimensional components are
given through the following expressions
\begin{equation}
T_{mn}=\frac{3}{8\pi G_5} e^{2A}\left[A''+2(A')^2\right]\eta_{mn},
\qquad T_{55}=\frac{6(A')^2}{8\pi G_5}.
\end{equation}

In order to find a solution to the setup defined by (\ref{action})
and (\ref{line}), we shall use the conformal technique: by means of
a conformal transformation we go from the Weyl frame to the Riemann
one, where the Weylian affine connections become Christoffel symbols
and the field equations are simpler, solve these equations and
return to the Weyl frame to analyze the physics of the solution. By
performing the conformal transformation
$\widehat{g_{MN}}=e^{\omega}g_{MN}$ the action (\ref{action}) is
mapped into a Riemannian manifold
\begin{equation}
\label{confaction} S_5^R=\int_{M_5^R}\frac{d^5x\sqrt{|\widehat
g|}}{16\pi G_5}\left[\widehat
R+3{\xi}\left(\widehat\nabla\omega\right)^2+6\widehat
U(\omega)\right],
\end{equation}
where $\xi=\tilde{\xi}-1$, $\ \widehat U(\omega)=e^{-\omega}
U(\omega)$ and all the hatted magnitudes and operators refer to the
Riemann frame. Thus, in this frame, we have a theory which describes
5--dimensional gravity coupled to a scalar field with a
self--interaction potential. After this transformation, the line
element is given by the following expression
\begin{equation}
\label{conflinee} \widehat{ds}_5^2=e^{2\sigma(y)}\eta_{nm}dx^n
dx^m+e^{\omega(y)}dy^2,
\end{equation}
where $2\sigma=2A+\omega$. Now the expressions for the Ricci tensor
and the scalar curvature in the Riemann frame read
\begin{equation}
\widehat{R}_{mn}=-e^{2A}\left[\sigma''+4(\sigma')^2-\frac{1}{2}\omega'\sigma'\right]\eta_{mn},\qquad
\widehat{R}_{55}=-4\left[\sigma''+(\sigma')^2-\frac{1}{2}\omega'\sigma'\right],\nonumber
\end{equation}
\begin{equation}
\widehat{R}=-4e^{-\omega}\left[2\sigma''+3(\sigma')^2+2\sigma'A'\right].
\end{equation}
The 4--dimensional and pure 5--dimensional components of the
stress--energy tensor are given by
\begin{equation}
\widehat{T}_{mn}=\frac{3}{8\pi G_5}
e^{2A}\left[\sigma''+2(\sigma')^2-\frac{1}{2}\omega'\sigma'\right]\eta_{mn},
\qquad \widehat{T}_{55}=\frac{6(\sigma')^2}{8\pi G_5}.
\end{equation}

\paragraph{The solution.} In order to find a solution for the model a new
pair of variables $X\equiv\omega'$ and $Y\equiv2A'$ are introduced.
Thus, the corresponding to (\ref{confaction}) field equations with
the scaled line element (\ref{conflinee}) read \cite{ariasetal}
\begin{eqnarray}
\label{fielde}
X'+2YX-\frac{3}{2}X^2&=&\frac{1}{\xi}\frac{d\widehat U}{d\omega}e^{-\omega},\nonumber\\
Y'+2Y^2-\frac{3}{2}XY&=&\left(\frac{1}{\xi}\frac{d\widehat
U}{d\omega}+4\widehat U\right)e^{-\omega}.
\end{eqnarray}

If we assume the condition $X=kY$, where $k$ is an arbitrary
constant this system of equations can be easily solved. Under this
restriction, the potential $\widehat U$ must have the following form
$\widehat U=\lambda e^{\frac{4k\xi}{1-k}\omega}$.

After imposing these conditions, the equation system (\ref{fielde})
becomes
\begin{equation}
\label{finale}
Y'+\frac{4-3k}{2}Y^2=\frac{4\lambda}{1-k}e^{(\frac{4k\xi}{1-k}-1)\omega}.
\end{equation}

By setting $\xi=\frac{1-k}{4k}$, the field equation (\ref{finale})
simplifies further to
\begin{equation}
\label{simfielde} Y'+\frac{4-3k}{2}Y^2=\frac{4\lambda}{1-k}.
\end{equation}
This choice accounts to having a self--interaction potential of the
form $U=\lambda e^{2\omega}$ in the Weyl frame which, indeed, does
not preserve the invariance of (\ref{action}) under Weyl rescalings.
By taking into account that $Y\equiv 2A'$ we get
\begin{equation}
\label{A''} A''+(4-3k)(A')^2-\frac{2\lambda}{1-k}=0.
\end{equation}
By solving (\ref{A''}) we find the following solution
\begin{eqnarray}
\label{pairsolut}
e^{2A(y)}=k_3\left(e^{ay}+k_1e^{-ay}\right)^b,\quad\quad
\omega=\ln\left[k_2\left(e^{ay}+k_1e^{-ay}\right)^{kb}\right],
\end{eqnarray}
where
\begin{eqnarray}
a=\sqrt{\frac{4-3k}{1-k}2\lambda}, \qquad \qquad \qquad
b=\frac{2}{4-3k},
\end{eqnarray}
and $k_1$, $k_2$ and $k_3$ are arbitrary constants.

This represents a solution which does not respect $Z_2$--symmetry
($y\longrightarrow -y$) due to the presence of the constant
parameter $k_1$. If we look at the particular case when $k_1=1$,
$k_2=2^{-kb}$ and $k_3=2^{-b}$ we recover the $Z_2$--symmetric
solution previously found by \cite{ariasetal} in the Weyl frame and
by \cite{dewolfe}--\cite{gremm} in the Riemann one:
\begin{eqnarray}
\label{solariasetal} e^{2A(y)}=\left[\cosh(ay)\right]^b,\quad\quad
\omega=bk\ln \left[\cosh(ay)\right].
\end{eqnarray}
In what follows we set $k_2=k_3=1$ since these constants are not
physically meaningful; they can be absorbed by a coordinate
redefinition in the metric (\ref{conflinee}).

\section{Physics of the system}

Let us turn to study the physics of the found solution in both the
Einstein and the Weyl frames. Since the conformal technique is used
to generate the solution, we first analyze the behaviour of the warp
factor, the energy density of the scalar field, the curvature scalar
of the system as well as the metric fluctuations in the Riemann
frame and after that we consider our solution and perform the same
analysis in the Weyl frame. We finally compare both results.

\paragraph{The Riemann frame.} The warp factor and the scalar field are
given by equations (\ref{pairsolut}) and their shape depends on the
values of the constants $k_1, a, b$. Thus, we have a family of
solutions depending on the values of the parameters $k_1$, $\lambda$
and $k$. We look for cases where the warp factor is a localized,
smooth and well behaved function which in fact models the fifth
dimension. By taking into account this fact, we have the following
cases of interest:

A) $k_1>0$, $\lambda>0$ and $k>\frac{4}{3}$. In this case we have
$a\in\Re$, $b<0$, while the range of the fifth coordinate is
$-\infty<y<\infty$, thus, we have a non--compact manifold in the
extra dimension. It turns out that in this case, both the warp
factor $e^{2A}$ and $e^{\omega}$ are symmetric functions with
respect to the point $y_0=\ln k_1/(2a)$ and represent smooth, well
behaved but non $Z_2$--symmetric localized functions characterized
by the width parameter $\Delta\sim 1/a$ (see Fig. 1).

\EPSFIGURE[pos]{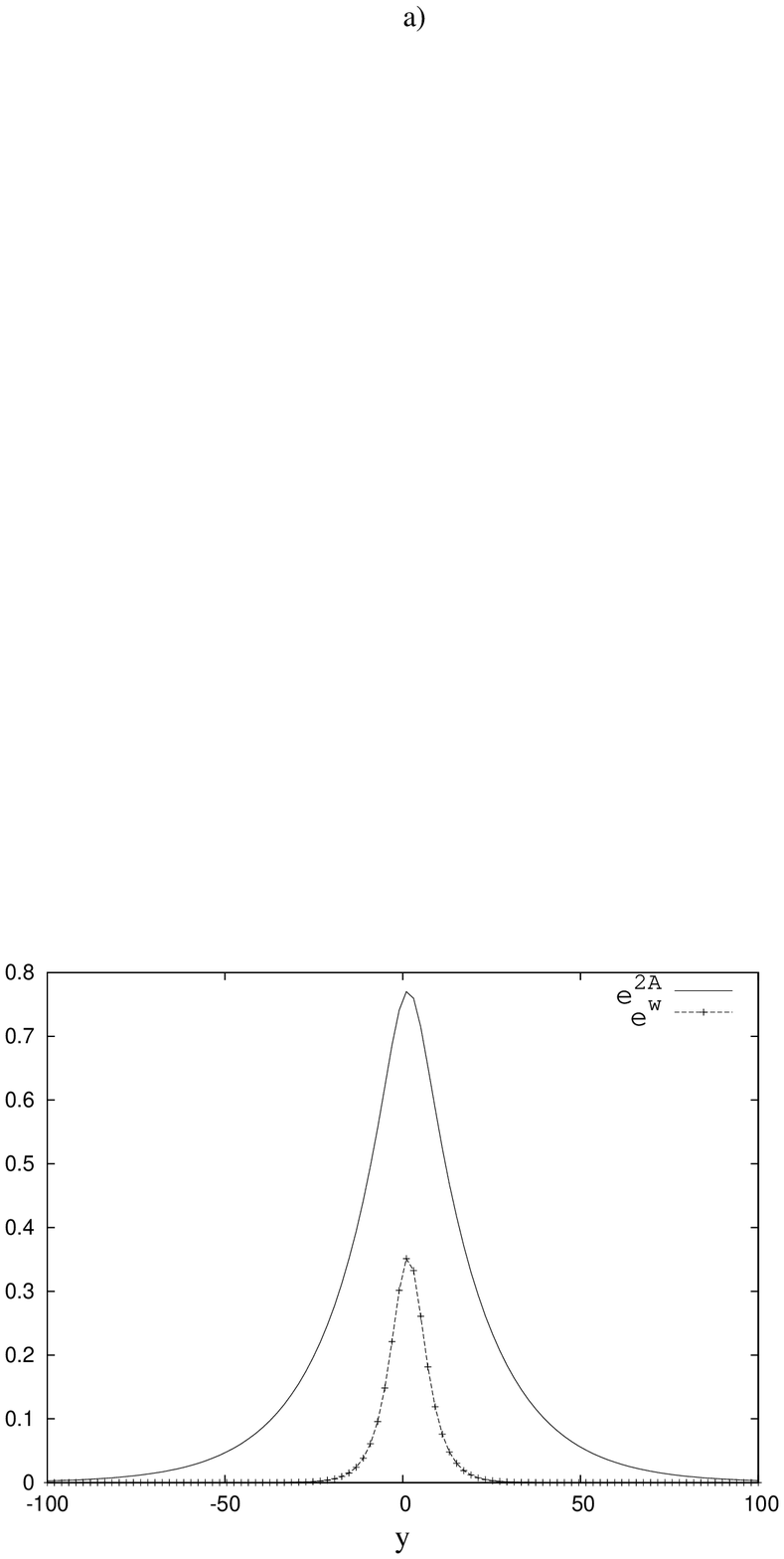,width=8cm}{\label{fig:warp}The
shape of the warp factor $e^{2A}$ and $e^{\omega}$ for $k>4/3$,
$\lambda>0$, and $k_1>0$ is a smooth, localized and non
$Z_2$--symmetric function. In this particular figure we have set
$k=4$, $\lambda=1/100$, and $k_1=2$ for both frames.}

\noindent It is easy to see that $k_1$ measures the $Z_2$--asymmetry
of the solution through a shift along the positive axis of the extra
coordinate given by the value $y_0$. Since this constant appears
multiplying an exponential function of $y$, its effect is quite
small and, hence, the solution slightly deviates from the
$Z_2$--symme\-tric one. However, the physical implications of this
fact are quite important, namely, the 5--di\-men\-sio\-nal space
time is not restricted to be an orbifold geometry, allowing for a
more general kind of dimensional reductions when going down to four
dimensions.

\noindent The 4--dimensional and pure 5--di\-men\-sio\-nal
components of the Riemannian stress--energy tensor are given by the
following expressions
\begin{equation}
\widehat {T}_{mn}=\frac{6a^2b(1+k)k_1}{8\pi
G_5}(e^{ay}+k_1e^{-ay})^{b-2}\left[1+\frac{b(k+2)}{8k_1}(e^{ay}-k_1e^{-ay})^2
\right]\eta_{mn}
\end{equation}
and
\begin{equation}
\widehat {T}_{55}=\frac{3a^2b^2(1+k)^2}{16\pi
G_5}\left(\frac{e^{ay}-k_1e^{-ay}}{e^{ay}+k_1e^{-ay}}\right)^2.
\end{equation}
The energy density of the scalar matter is given by the null-null
component of the stress-energy tensor
\begin{equation}
\widehat{\mu}(y)=\frac{-6a^2b(1+k)k_1}{8\pi
G_5}(e^{ay}+k_1e^{-ay})^{b-2}\left[1+\frac{b(k+2)}{8k_1}(e^{ay}-k_1e^{-ay})^2
\right].
\end{equation}

The shape of the energy density $\widehat\mu$ shows a positive
maximum at $y_0$ and two negative minima around the maximum; the
function vanishes asymptotically for $y=\pm\infty$. Thus, this
configuration can be interpreted as a thick brane with positive
energy density, where the scalar matter is confined, located between
two thick branes with negative energy density (see Fig. 2). This
thick brane configuration becomes $Z_2$--symmetric in the particular
case when $k_1=1$, resembling the thin brane Randall--Sundrum model
\cite{rs}.

\EPSFIGURE[pos]{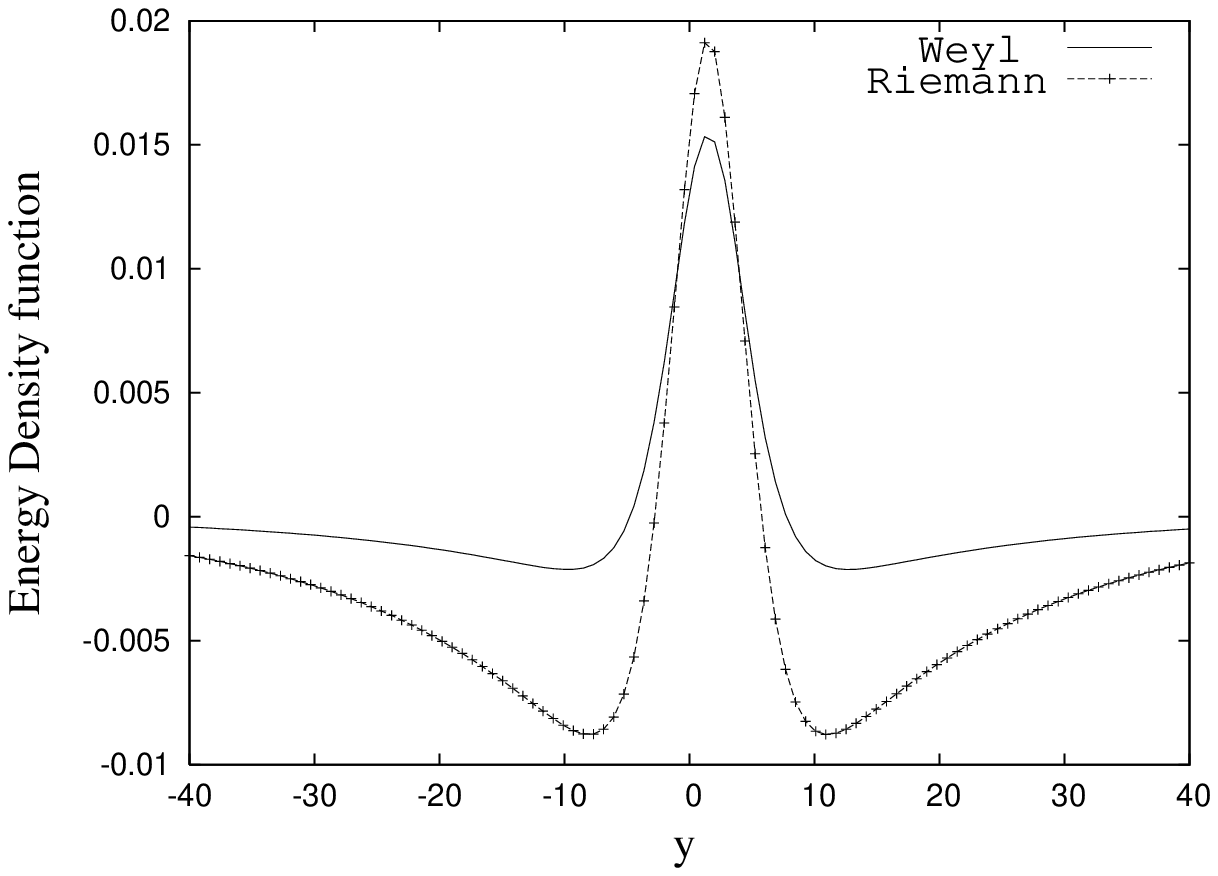,width=8cm}{\label{fig:desityw}The
shape of the energy density function with the same parameter values
as in Fig. 1, but we have rescaled $\widehat\mu$ by 1/4. Two thick
branes with negative density energy are located around a brane with
positive energy density centered at $y_0=\ln k_1/(2a)$.}

The 5--dimensional curvature scalar
\begin{eqnarray}
\widehat R_5&=&\frac{-64\lambda k_1(1+k)}{1-k}(e^{ay}+k_1e^{-ay})^{-(kb+2)}\nonumber\\
&\times&\left[1+\frac{b(5+3k)}{16k_1}(e^{ay}-k_1e^{-ay})^2\right]
\end{eqnarray}
is not bounded and we have a 5--dimen\-sio\-nal manifold which is
singular in the Riemann frame.

B) $k_1>0$, $\lambda>0$ and $1<k<4/3$. In this case we have
$a\in\Im$, $b>0$ and we must replace $a\rightarrow i\alpha$ in order
to have a real warp factor. Moreover, the only possible choice for
the parameter $k_1$ is $k_1=1$ (otherwise the solution becomes
complex), and we get a $Z_2$--symmetric function
\begin{equation}
e^{2A(y)}=\cos^b(\alpha y).
\end{equation}
Thus, this represents a manifold which is periodic in the extra
dimension, so $-\pi\le \alpha y \le\pi$, and we have the same
compact case that was obtained in \cite{ariasetal}.

The physically relevant case $\lambda<0$, $k<1$, $k_1>0$ coincides
with case B), while the cases $\lambda>0$, $1<k<4/3$, $k_1<0$ and
$\lambda<0$, $k<1$, $k_1<0$ yield $y$--periodic warp factors
proportional to $\sin^b(ay)$ for positive $b$. These cases also
correspond in essence to the case B) shifted along the fifth
dimension. All other possible combinations of the parameters do not
yield physically relevant cases since for them, the warp factor
infinitely increases as $y$ approaches certain values and thus, it
does not constitute a localized function.

\paragraph{Metric Fluctuations.} In order to see whether we can describe
4--dimensional gravity within our setup, we will examine
fluctuations of the metric around a classical background solution.
Thus, the perturbed metric (\ref{conflinee}) reads
\begin{equation}
\widehat{ds}^2=e^{2\sigma (y)}[\eta_{mn}+\widehat h_{mn}(x,y)]dx^m
dx^n+ e^{\omega (y)}dy^2.
\end{equation}

In general, it is not possible to avoid considering fluctuations of
the scalar field while treating fluctuations of the background
metric. However, there exists a sector of the metric that decouples
from the scalar field and the metric fluctuations can be suited
analytically \cite{dewolfe}.

Following this line, we introduce a new $r$ coordinate $dr=e^{-A}dy$
which yields a conformally flat metric
\begin{equation}
\widehat{ds}^2=e^{2\sigma
(y)}\left\{[\eta_{mn}+\widehat{h}_{mn}(x,y)]dx^m dx^n+dr^2\right\}
\end{equation}
and the corresponding wave equation for the transverse traceless
modes $\widehat h_{mn}^T$ that decouple from the scalar field
\begin{equation}
\label{mttn}
\left(\partial_r^2+3\sigma'\partial_r+\Box^\eta\right)\widehat
h_{mn}^T=0,
\end{equation}
where $\Box^\eta$ is the (flat) Minkowski wave operator. In
\cite{dewolfe} it has been shown that this equation supports a
naturally massless and normalizable 4--dimensional graviton, since
it is massless, it must obey $\Box^\eta\widehat h_{mn}^T=0$, and we
can easily see that such a solution is given by the following ansatz
\begin{equation}
\widehat h_{mn}^T=C_{mn}e^{imx},
\end{equation}
where the $C_{mn}$ are constants and $m^2=0$.

For studying the massive KK--excitations it is useful to convert
equation (\ref{mttn}) into a Schrodinger's equation form. In  order
to achieve this aim we make the ansatz
$\widehat{h}_{mn}^T=e^{imx}e^{-3\sigma/2}\Psi_{mn}(r) $ and the
equations simplify further to
\begin{equation}
\label{schrodingerriem} [\partial_r^2-\widehat V(r)+m^2]\Psi=0,
\end{equation}
where we have dropped the subscripts in $\Psi$, $m$ is the mass of
the KK--excitation and the potential $\widehat V(r)$ reads
\begin{equation}
\widehat
V(r)=\frac{3}{2}\partial_r^2\sigma+\frac{9}{4}(\partial_r\sigma)^2.
\end{equation}
The shape of the potential $\widehat V(r)$ for the case A) of the
solution (\ref{pairsolut}) is given by the following expression:
\begin{equation}
\label{potentialrem} \widehat V(r(y))=
3a^2b(k+1)k_1(e^{ay}+k_1e^{-ay})^{b-2}\Big[1+\frac
{(3k+5)b}{16k_1}(e^{ay}-k_1e^{-ay})^2\Big].
\end{equation}

By looking at Fig. 3 we see that this potential represents a well
with a finite negative minimum at a certain value $y_0$, which is
located between two positive maxima (potential barriers) and then
vanishes as $y=\pm\infty$ (a volcano potential with finite bottom).

In the particular case $k=5/3$ (hence $b=-2$), the coordinate
transformation can be successfully inverted and we get
$\sigma(r)=-4\ln\left[\left(a^2r^2+4k_1\right)\right]/3$ which
yields a potential of the form
\begin{equation}
\label{potentialr} \widehat V(r)=
\frac{4a^2\left(5a^2r^2-4k_1\right)}{\left(a^2r^2+4k_1\right)^2}.
\end{equation}

\EPSFIGURE[pos]{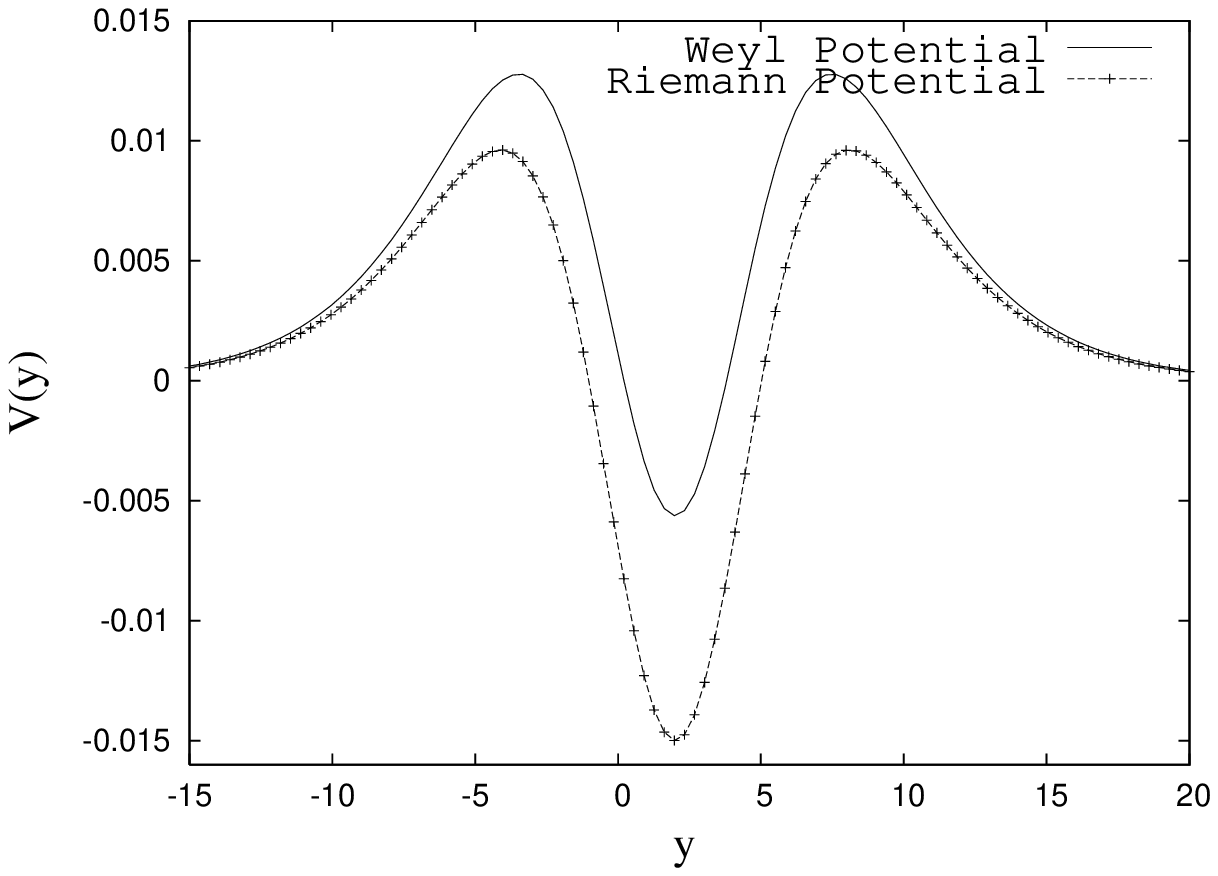,width=8cm}{\label{fig:potential}The
shape of the potential $V(y)$. For this figure we have set $k=5/3$,
$\lambda=1/100$, and $k_1=2$ in both frames.}

The spectrum of eigenvalues $m^2$ parameterizes the spectrum of
graviton masses that a 4--dimensional observer located at $y_0$
sees. It turns out that for the zero mode $m^2=0$, the Schr\"odinger
equation (\ref{schrodingerriem}) can be solved. The only
normalizable eigenfunction reads
$\Psi_0=c\left(a^2r^2+4k_1\right)^{-2}$, where $c$ is a
normalization constant. Since this function has no zeros, it
represents the lowest energy eigenfunction of the Sch\"odinger
equation (\ref{schrodingerriem}), allowing for the existence of a
4--dimensional graviton with no instabilities from transverse
traceless modes with  $m^2<0$; moreover, since the potential
vanishes asymptotically, this is the only gravitational bound state.
In addition there exists a tower of higher KK modes with positive
$m^2>0$. Thus, we have obtained non $Z_2$--symmetric thick brane
generalization of the Randall--Sundrum model in which the
4--dimensional effective theory possesses an energy spectrum quite
similar to the spectrum of the thin wall case. Similar results were
obtained in \cite{gremm}, but with just one free parameter.

\paragraph{The Weyl frame.} Once we have obtained the solution in the
Riemann frame by means of the conformal technique, we come back to
the Weyl frame to study its physics. In this frame, for the
non--compact case A) $\lambda>0$, $k>\frac{4}{3}$ and $k_1>0$, the
pure 5--dimensional and 4--dimensional components of the
stress--energy tensor are given through the following expressions
$$
T_{55}=\frac{3a^2b^2}{16\pi
G_5}\left(\frac{e^{ay}-k_1e^{-ay}}{e^{ay}+k_1e^{-ay}}\right)^2
$$
and
\begin{equation}
\label{TmnW} T_{mn}=\frac{6a^2bk_1}{8\pi
G_5}(e^{ay}+k_1e^{-ay})^{b-2}
\left[1+\frac{b}{4k_1}(e^{ay}-k_1e^{-ay})^2 \right]\eta_{mn}.
\end{equation}

\noindent Thus, the energy density of the scalar matter is given by
\begin{equation}
\mu(y)=\frac{-6a^2bk_1}{8\pi
G_5}(e^{ay}+k_1e^{-ay})^{b-2}\left[1+\frac{b}{4k_1}(e^{ay}-k_1e^{-ay})^2
\right].
\end{equation}

This function also shows two negative minima and a positive maximum
between them and, finally, it vanishes asymptotically. Thus, it
presents the same behaviour as the energy density in the Riemann
frame. It is displayed in Fig. 3, together with the same Riemannian
quantity $\widehat\mu$ to facilitate their comparison.

The 5-dimensional curvature scalar reads
\begin{equation}
\label{R5W} R_5= \frac{-16a^2bk_1}{(e^{ay}+k_1e^{-ay})^2}
\left[1+\frac{5b}{16k_1}(e^{ay}-k_1e^{-ay})^2 \right].
\end{equation}

From Fig. 4 we can see that this quantity is always bounded, thus,
in contrast with the singular manifold that arises in the Riemann
frame, we have a 5--dimensional manifold that is regular in the Weyl
frame\footnote{We thank the authors of \cite{ariasetal} for
clarifying the discrepancy between some of their expressions and
ours in the limit $k_1=1$; they kindly communicated us that they
made a systematic mistake when computing the quantities
corresponding to (\ref{TmnW})--(\ref{R5W}) in their work.}.

\EPSFIGURE[pos]{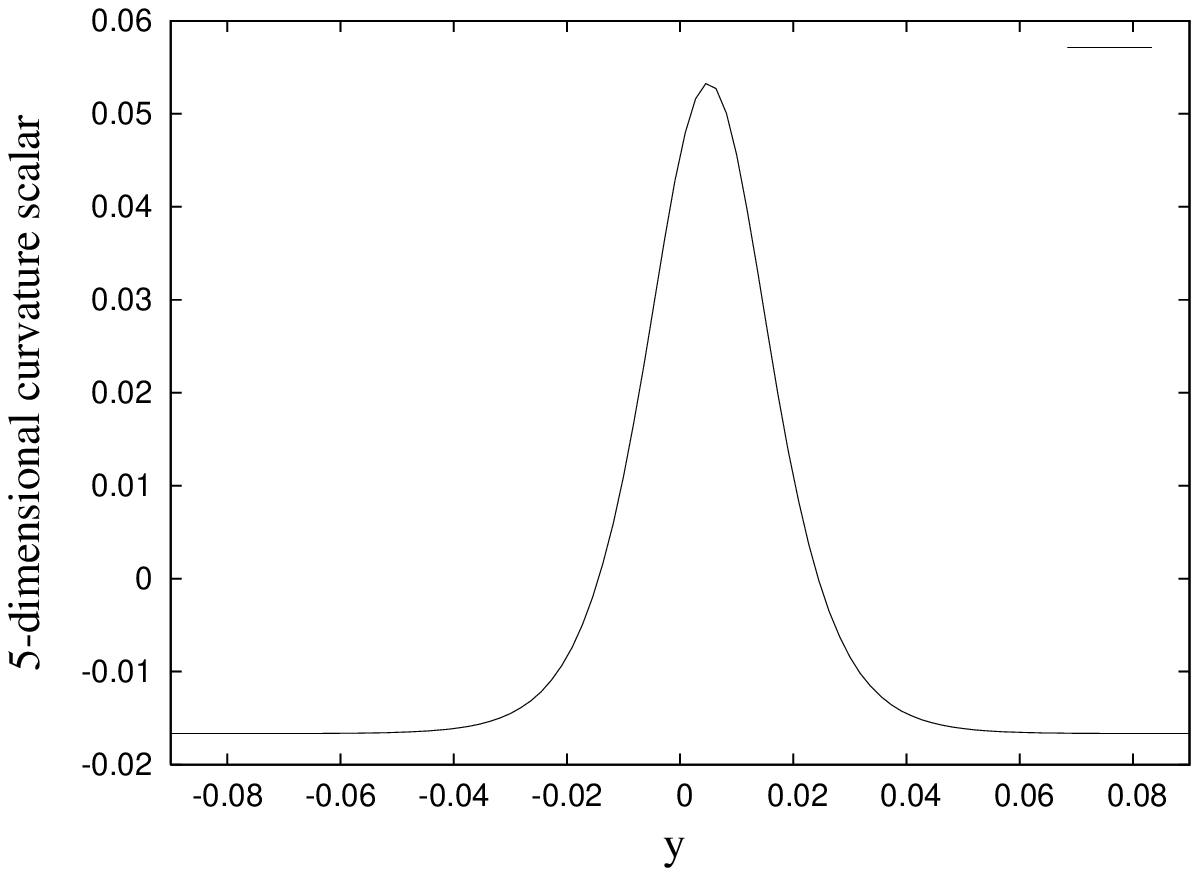,width=8cm}{\label{fig:curvw}The
functional form of the curvature scalar $R_5$ for $k>4/3$ in the
Weyl frame. This magnitude is singular in the Riemann frame.}

{\bf Metric Fluctuations.} Now we want to study the metric
fluctuations $h_{mn}$ in the Weyl frame. In order to achieve this
goal, we consider perturbations of the metric (\ref{line})
\begin{equation}
\label{mfluct} ds_5^2=e^{2 A(y)}[\eta_{mn}+h_{mn}(x,y)]dx^m
dx^n+dy^2
\end{equation}
and perform the coordinate transformation $dw=e^{-A}dy$ which leads
to a conformally flat metric and to the following wave equation for
the transverse traceless modes $h_{mn}^T$ of the metric fluctuations
\begin{equation}
\label{eqttm} (\partial_w^2+3A'\partial_w+\Box)h_{mn}^T=0.
\end{equation}
It is easy t see that this equation supports as well a massless and
normalizable 4--dimensional graviton in the Weyl frame given by
$h_{mn}^T= K_{mn}e^{imx}$, where $K_{mn}$ are constants and $m^2=0$.

\noindent In order to recast equation (\ref{eqttm}) into a
Schr\"{o}dinger's equation form we adopt the ansatz
$h_{mn}^T=e^{imx}e^{-3A/2}\psi_{mn}(w)$, and equation (\ref{eqttm})
simplifies further to
\begin{equation}
\label{schrodinger} [\partial_w^2-V(w)+m^2]\psi=0,
\end{equation}
where we have dropped the subscripts in $\psi$, $m$ is the mass of
the KK--excitation, and the potential reads
\begin{equation}
\label{VW} V(w)=\frac{3}{2}\partial_w^2A+\frac{9}{4}(\partial_wA)^2.
\end{equation}

The shape of the potential $V(w)$ in this set-up is given by the
following expression
\begin{equation}
\label{potential} V(w(y))=3a^2bk_1
(e^{ay}+k_1e^{-ay})^{b-2}\Big[1+\frac
{5b}{16k_1}(e^{ay}-k_1e^{-ay})^2\Big]
\end{equation}
and its behaviour is displayed in Fig. 3. From (\ref{potential}) we
observe that it possesses a similar structure in comparison to its
Riemannian analogue; this is also confirmed by the shape of the
potential in Fig. 3.

Let us consider the case $k=5/3$ ($b=-2$) in which the
coordinate transformation $dw=e^{-A}dy$ can be inverted. The
function $A$ reads $A(w)=-\ln\left(a^2w^2+4k_1\right)/2$ and yields
a potential of the form
\begin{equation}
\label{potentialr} V(r)=
\frac{3a^2\left(5a^2w^2-8k_1\right)}{4\left(a^2w^2+4k_1\right)^2}.\end{equation}

Once again, the Schr\"odinger equation (\ref{schrodinger}) can be
solved for the massless zero mode $m^2=0$, leading to the only
normalizable eigenfunction $\psi_0=q\left(a^2w^2+4k_1\right)^{3/4}$,
where $q$ is a normalization constant, which represents the
4--dimensional graviton with no instabilities from transverse
traceless modes with  $m^2<0$. Thus, the spectrum of graviton masses
that a 4--dimensional observer located at $y_0$ sees in the Weyl
frame is qualitatively the same as that of the Riemann frame and we
have as well a continuum spectrum of KK states with positive
$m^2>0$.

This fact contradicts the result obtained in \cite{ariasetal} where
a discrete spectrum of massive KK modes was obtained. It seems to us
that this is a consequence of the form of their potential (see
equation (26) in that work), where the power of the hyperbolic
cosine function is the opposite to ours. Let us clarify this point.
By setting $k_1=1$ in our solution we get the solution obtained by
Arias et al. Further, by substituting the expression for $A(y)$ in
the formula for the potential (\ref{VW}), we get an expression
proportional to the warp factor $\cosh^b(ay)$ multiplied by
$[1+\beta\tanh^2(ay)]$, where $\beta$ is a constant. One can see
that both factors are bounded magnitudes for any real $a$ and $b<0$
(this is precisely the particular non--compact case we are
considering). Thus, there is no way to obtain a potential well with
infinite rigid walls and, therefore, the the spectrum of KK states
turns out to be continuum and gapless. Thus, we conclude that for
the setup considered in the present work, Weyl geometries do not
mimic classically quantum behaviour.

\section{Conclusions and Discussion}

Along with the study of the physics of our thick brane solution in a
5--dimensional Riemannian space time we have also considered it in a
geometric generalization of Kaluza--Klein theory obtained by
replacing the Riemannian structure of the 5--dimensional manifold
with a pure geometric Weyl integrable manifold in which variations
in the length of vectors during parallel transport are allowed and a
geometric scalar field is involved in its formulation. This was
motivated by the hypothesis that Weyl geometries mimic quantum
behaviour classically \cite{ariasetal}. However, as a result of our
investigation, we have shown that this is not a generic feature of
such manifolds.

By means of the conformal technique we obtained a classical
4--dimensional Poincar\'e invariant solution which represents a well
behaved localized function (a thick brane characterized by the width
parameter $\Delta\sim 1/a$) which does not respect $Z_2$--symmetry
along the extra dimension. Thus, we have obtained a thick brane
generalization of the Randall--sundrum model. This field
configuration does not restrict the 5--dimensional space time to be
an orbifold geometry, allowing for arbitrary positive and negative
values along the fifth coordinate $y$. This fact can be exploited
when addressing difficult cosmological issues such as the
cosmological constant problem, black hole physics \cite{lr} and
holographic ideas, where there exists a correspondence between
location in the fifth dimension and mass scale \cite{verlinde},
since one has a better chance of solving these problems when all
possible mass scales (all possible distances along the fifth
dimension) are available. When we set the parameter $k_1=1$, our
solution reproduces the $Z_2$--symmetric solutions previously found
in the literature in both the Riemann and the Weyl frames.

By looking at the scalar energy density $\mu$ of our field
configuration, we see that it shows a thick brane with positive
energy density centered at $y_0$ and accompanied by one thick brane
with negative energy density at each side. The fact that we have
both negative and positive values for the scalar energy density can
be compared with the standard Randall--Sundrum thin brane case. The
generic behaviour of this physical quantity is the same in both the
Riemann and the Weyl frame.

The scalar curvature of the Riemannian manifold turns out to be
singular for the found solution, whereas the corresponding quantity
in the Weyl integrable geometry presents a regular behaviour along
the whole fifth dimension. It is interesting to check whether this
is a generic property of the considered models.

By studying the behaviour of the transverse traceless modes of the
fluctuations of the metric we recast their equations into a one
dimensional Sch\"odinger equation with a quantum mechanical
potential that represents a finite negative well with a finite
positive barrier at each side which then vanishes asymptotically (a
volcano potential with finite bottom). We solve the Sch\"odinger
equation for the massless zero mode $m^2=0$ in both frames obtaining
a single bound state which represents the lowest energy
eigenfunction of the Sch\"odinger equation, allowing for the
existence of a 4--dimensional graviton with no instabilities from
transverse traceless modes with $m^2<0$. We also get a huge tower of
higher KK states with positive $m^2>0$ that are suppressed at $y_0$,
turning into continuum plane wave modes as $y$ approaches spatial
infinity \cite{lr}--\cite{dewolfe}. Since all values of $m^2$ are
allowed, the spectrum turns out to be continuum and gapless.

We prove that for our setup, the analog quantum mechanical potential
does not possess the shape of a well with infinite rigid walls. Due
to this fact, one cannot get a discrete (quantized) spectrum of KK
states when analyzing fluctuations of the metric. Thus, we conclude
that the claim that Weylian structures mimic, classically, quantum
behaviour \cite{ariasetal}, \cite{quantbeh} does not constitute a
generic feature of these geometric manifolds.

Due to the dependence of the quantum mechanics potential on the warp
factor functions, different warp factors will lead to distinct
potentials and these, in turn, will alter the Schr\"odinger
equation. When solving it, the new solutions will describe wave
functions with different physics. Thus, it is important to consider
other solutions to the original setup (both in Einstein and Weyl
frames) and analyze the physics they yield. Another interesting
issue that is currently under investigation is the solution of the
mass hierarchy problem in 5D space times involving thick brane
configurations.

\bigskip

\acknowledgments

The authors are really grateful to O. Arias and I. Quiros for a
detailed explanation of the conformal technique and for stimulating
correspondence on the subject. They thank as well U. Nucamendi for
useful discussions while this investigation was carried out. One
author (AHA) would like to express his gratitude to the Theoretical
Physics Department of the Aristotle University of Thessaloniki and,
specially, to Prof. J.E. Paschalis for fruitful discussions and for
providing a stimulating atmosphere while this work was in progress.
This research was supported by grants CIC-UMSNH-4.16 and
CONACYT-F42064.

\listoffigures          

\end{document}